\documentclass[aps,groupedaddress,showpacs,reprint,dvipdfmx]{revtex4-2}
\usepackage{graphicx}
\usepackage{color}
\usepackage{amsmath}
\usepackage{bm}
\usepackage{ulem}

\begin{document}

\title{
Irreversibility of the pendulum revisited from Husimi's adiabaticity parameter
}

\author{Yuki Izumida}
\affiliation{Department of Complexity Science and Engineering, Graduate School of Frontier Sciences, The University of Tokyo, Kashiwa 277-8561, Japan}
\thanks{izumida@k.u-tokyo.ac.jp}

\begin{abstract}
We revisit the irreversibility of the pendulum with time-dependent angular frequency, considered in a classical paper by K. Husimi. 
He introduced a parameter that measures the adiabaticity of a process utilizing an adiabatic invariant for the equation of motion of the pendulum.
With this adiabaticity parameter, Husimi showed the irreversibility of the pendulum for a cyclic process, which is reminiscent of the Planck principle in thermodynamics, based on the microscopic mechanics.
In this study, we generalize the argument by Husimi to a damped pendulum with friction, and highlight the role of conservation of a phase-space area 
on the Husimi's adiabaticity parameter.
Moreover, we also investigate the second law of thermodynamics and its generalization for a general non-cyclic process as well as a cyclic process, and elucidate how the Husimi's adiabaticity parameter impacts on this law. In particular, for a general non-cyclic process, we show the law of entropy non-decrease for the pendulum without friction
by using the property of the Husimi's adiabaticity parameter.
\end{abstract}

\maketitle

\section{Introduction}
Deriving irreversibility from an underlying microscopic mechanics is one of the goals in nonequilibrium statistical mechanics since Boltzmann.
Such attempts have been made for both classical and quantum mechanical systems, and under some reasonable assumptions the second law of thermodynamics 
for time-dependent operations were derived in general setups~\cite{LA1978,T2000,SK2000,AN2005,C2008_1,C2008_2,C1997,C2020,S2020}.

One of the key points to connect thermodynamics and mechanics is to focus on an adiabatic theorem in mechanics.
In classical mechanics, when a parameter of a system is adiabatically (sufficiently slowly) changed, the phase-space volume enclosed by a constant energy surface is proportional to an adiabatic invariant and is conserved in time.
Because the thermodynamic entropy in macroscopic systems is given by the logarithm of the phase-space volume enclosed by a constant energy surface, 
its conservation may correspond to the conservation of the entropy of a thermally-isolated system under a quasistatic operation~\cite{footnote1}.

In a classical paper~\cite{H1953}, Husimi established a connection between thermodynamics and mechanics in terms of the adiabatic theorem.
In an attempt to calculate a transition probability of a quantum parametric oscillator, Husimi found that the 
generating function of the transition probability is characterized by an adiabaticity parameter. 
It measures the adiabaticity of a process in terms of an adiabatic invariant of the corresponding {\it classical} parametric oscillator, which we refer to as the pendulum hereafter.
The parameter takes unity under an adiabatic change of a parameter reflecting the existence of adiabatic invariants while it is larger than unity under a non-adiabatic change, thus serving as a degree of adiabaticity.
By using this property of the adiabaticity parameter, Husimi succeeded in showing the thermodynamic irreversibility of the pendulum in a cyclic operation. 
To be specific, he proved the following relation:
\begin{eqnarray}
E_{t_1}=E_{t_0}Q_{t_1}^*.
\end{eqnarray}
Here, $E_{t_0}$ and $E_{t_1}$ are the average energies of the pendulum before and after the cyclic operation starting from an initial equilibrium distribution, and $Q_{t_1}^*$ is the Husimi's adiabaticity parameter (see Eq.~(\ref{eq.Q_def}) for its definition).
Because of the property $Q_{t_1}^* \ge 1$, we obtain
\begin{eqnarray}
E_{t_1}\ge E_{t_0}.\label{eq.Planck_intro}
\end{eqnarray}
This is reminiscent of the Planck principle in thermodynamics, which was derived based on mechanics.
For an interpretation of this inequality, let us quote from the Husimi's paper (p.~391 of~\cite{H1953}):

\vspace{5pt}
``Let the string of a pendulum, while in oscillation, be shortened or lengthened in an arbitrary manner and let it return to its initial length after a time. Then we expect,
{\it on the average}, to find the energy of the pendulum increased."
\vspace{5pt}

The Husimi's adiabaticity parameter received renewed interests in thermodynamics of small systems and found many applications such as in work distributions~\cite{DL2008,SD2021,JDG2017}, finite-time heat engines~\cite{ARJDSKSL2012,RASKSL2014,JBdC2016}, and shortcuts to adiabaticity~\cite{BJdC2016,MI2017,AL2018,AP2019,SMDB2022}.

In this paper, we revisit the irreversibility of the pendulum Husimi studied in~\cite{H1953}, and generalize it.
Our generalizations are twofold: First, we generalize the setup so that it includes a damped pendulum with friction.
This generalization highlights the role of non-Hamilton dynamics as underlying mechanics on the Husimi's adiabaticity parameter.
Such a generalization would also be of interest in the context of thermodynamics for non-Hamilton systems~\cite{MD2016}.
Second, we investigate the second law of thermodynamics and its generalization not only for a cyclic process but also for a non-cyclic process in terms of the Husimi's adiabaticity parameter, where these two correspond to the Planck's principle and the law of entropy non-decrease in thermodynamics, respectively.

The reminder of the paper is organized as follows.
In Sec.~\ref{Model}, we introduce our pendulum model and specify the setup of the model.
In Sec.~\ref{Main results}, we present our main results.
In Sec.~\ref{Concluding remarks}, concluding remarks are given.

\section{Model}\label{Model}
Let us consider a one-dimensional pendulum obeying the following equation of motion:
\begin{eqnarray}
\ddot z_t=-\omega_t^2 z_t-\Gamma_t \dot z_t,\label{eq.motion}
\end{eqnarray}
where the dot denotes time derivative, and $\omega_t >0$ and $\Gamma_t$ denote a time-dependent angular frequency and a time-dependent friction coefficient of the pendulum, respectively.
Here, we allow $\Gamma_t$ to take any value including a negative value so that the energy can be injected into the pendulum.
The case $\Gamma_t=0$ was considered in~\cite{H1953}, and the present setup thus generalizes it by including the non-vanishing cases $\Gamma_t \ne 0$.

For the statistical mechanics description, we introduce the momentum $p_t\equiv \dot z_t$ and write the equation of motion Eq.~(\ref{eq.motion}) as the following equation for $(z_t, p_t)$ in the phase plane:
\begin{eqnarray}
&&\dot z_t=p_t,\\
&&\dot p_t=-\omega_t^2z_t-\Gamma_t p_t,
\end{eqnarray}
which, for $\Gamma_t=0$, recovers the Hamilton dynamics with the Hamiltonian $H_t(z_t, p_t)$:
\begin{eqnarray}
H_t(z_t, p_t) \equiv \frac{p_t^2}{2}+\frac{\omega_t^2 z_t^2}{2}.\label{eq.E_def}
\end{eqnarray}
Under an adiabatic change of the angular frequency $\omega_t$, the following quantity is conserved as an adiabatic invariant of the pendulum:
\begin{eqnarray}
\frac{H_t(z_t, p_t)}{\omega_t}=\frac{p_t^2+\omega_t^2 z_t^2}{2\omega_t} \approx {\rm const.} \label{eq.action}
\end{eqnarray}

We introduce the average energy $E_t\equiv \left<H_t\right>$ that serves as the internal energy of the pendulum:
\begin{eqnarray}
E_t=\left<H_t\right>=\int_{-\infty}^{\infty} \int_{-\infty}^{\infty} dz_t dp_t f_t(z_t,p_t)H_t(z_t,p_t),\label{eq.E}
\end{eqnarray}
where $f_t(z_t,p_t)$ is the probability distribution at time $t$
and $\left<\cdot \right>$ denotes the average with respect to the probability distribution.
At an initial time $t=t_0$, the system is assumed to obey the microcanonical distribution of energy $E_{t_0}$ 
as the same initial distribution as the original consideration by Husimi~\cite{H1953}:
\begin{eqnarray}
f_{t_0}(z_{t_0},p_{t_0})&&=\frac{\delta \left(E_{t_0}-H_{t_0}(z_{t_0}, p_{t_0})\right)}{A_{t_0}},
\end{eqnarray}
where
\begin{eqnarray}
A_{t_0} &&\equiv \int_{-\infty}^{\infty}\int_{-\infty}^{\infty} dz_{t_0} dp_{t_0}\delta \left(E_{t_0}-H_{t_0}(z_{t_0}, p_{t_0})\right)\nonumber\\
&&=\frac{2\pi}{\omega_{t_0}}\label{eq.A0}
\end{eqnarray}
is a normalization factor.
The realization of this microcanonical ensemble is given as follows~\cite{S2002}. 
Let us make contact between the pendulum and a heat bath, and after a sufficient time, we detach the heat bath from the pendulum.
The pendulum may obey the canonical distribution, and we collect the states with energy $E_{t_0}$ sampled from this distribution, which constitutes the desired microcanonical ensemble.
We assume that no work cost is taken for this process.
For $t>t_0$, the dynamics obeys Eq.~(\ref{eq.motion}) and is deterministic.
The pendulum is thermally isolated during the time evolution for $\Gamma_t=0$.

\section{Main results}\label{Main results}
\subsection{Internal energy in terms of Husimi's adiabaticity parameter}
We calculate the internal energy in Eq.~(\ref{eq.E}) and expresses it in terms of Husimi's adiabaticity parameter.
The general solution to Eq.~(\ref{eq.motion}) can be written in terms of the fundamental solutions $X_t$ and $Y_t$:
\begin{eqnarray}
z_t=z_{t_0}Y_t+p_{t_0}X_t,\label{eq.z_XY}
\end{eqnarray}
where $X_t$ and $Y_t$ satisfy
\begin{eqnarray}
\ddot X_t=-\Gamma_t\dot X_t-\omega_t^2 X_t; \ X_{t_0}=0, \ \dot X_{t_0}=1,\label{eq.X}
\end{eqnarray}
and
\begin{eqnarray}
\ddot Y_t=-\Gamma_t\dot Y_t-\omega_t^2 Y_t; \ Y_{t_0}=1, \ \dot Y_{t_0}=0,\label{eq.Y}
\end{eqnarray}
respectively. Note that $X_t$ has dimension of time and $Y_t$ is dimensionless as seen from Eq.~(\ref{eq.z_XY}).
By differentiating Eq.~(\ref{eq.z_XY}) with respect to time $t$, we also have
\begin{eqnarray}
p_t=z_{t_0}\dot Y_t+p_{t_0}\dot X_t.\label{eq.dotz_XY}
\end{eqnarray}
Equations (\ref{eq.z_XY}) and (\ref{eq.dotz_XY}) relate $(z_t, p_t)$ and $(z_{t_0}, p_{t_0})$ in terms of the fundamental solutions $X_t$ and $Y_t$ and their time derivatives.

We can rewrite the internal energy in Eq.~(\ref{eq.E}) as
\begin{eqnarray}
E_t&&=\int_{-\infty}^{\infty}\int_{-\infty}^{\infty}dz_{t_0}d p_{t_0}\ \frac{\delta \left(E_{t_0}-H_{t_0}(z_{t_0}, p_{t_0})\right)}{A_{t_0}}\nonumber\\
&&\times H_t(z_t(z_{t_0}, p_{t_0}), p_t(z_{t_0}, p_{t_0})),\label{eq.E_int}
\end{eqnarray}
in terms of the integral with respect to $z_{t_0}$ and $p_{t_0}$, where we have used $f_t(z_t, p_t)dz_tdp_t=f_{t_0}(z_{t_0}, p_{t_0})dz_{t_0}dp_{t_0}$.
By putting Eqs.~(\ref{eq.z_XY}) and (\ref{eq.dotz_XY}) into Eq.~(\ref{eq.E_int}), and performing the integration with respect to $z_{t_0}$ and $p_{t_0}$, we obtain (see Appendix~\ref{U_deriv} for the derivation)
\begin{eqnarray}
E_t=E_{t_0} \frac{\omega_t}{\omega_{t_0}}Q_t^*.\label{eq.U}
\end{eqnarray}
Here, $Q_t^*$ is the Husimi's adiabaticity parameter defined as~\cite{H1953}:
\begin{eqnarray}
Q_t^*\equiv \omega_{t_0}\frac{\dot X_t^2+\omega_t^2X_t^2}{2\omega_t}+\omega_{t_0}^{-1}\frac{\dot Y_t^2+\omega_t^2 Y_t^2}{2\omega_t}. \label{eq.Q_def}
\end{eqnarray}
By considering the initial conditions in Eqs.~(\ref{eq.X}) and (\ref{eq.Y}), we have $Q^*_{t_0}=1$.

\subsection{Inequality for Husimi's adiabaticity parameter}
It can be shown that $Q_t^*$ is bounded from the lower side as (see Appendix~\ref{Husimi} for the derivation)
\begin{eqnarray}
Q_t^*\ge W_t, \label{eq.Q_inq}
\end{eqnarray}
where $W_t$ is the Wronskian for the fundamental solutions $X_t$ and $Y_t$ obeying Eqs.~(\ref{eq.X}) and (\ref{eq.Y}):
\begin{eqnarray}
W_t\equiv \dot X_tY_t-X_t\dot Y_t.\label{eq.def_W}
\end{eqnarray}
Note that
\begin{eqnarray}
W_{t_0}=1,\label{eq.W_ini}
\end{eqnarray}
from the initial conditions in Eqs.~(\ref{eq.X}) and (\ref{eq.Y}).
The Wronskian Eq.~(\ref{eq.def_W}) serves as an area expansion rate in the phase plane.
In fact, the time evolution of an infinitesimal area element from an initial time $t_0$ to a time $t$ in the phase plane is given by the Jacobian of the transformation from $(z_{t_0}, p_{t_0})$ to $(z_t, p_t)$:
\begin{eqnarray}
dz_tdp_t=|J_t| dz_{t_0}dp_{t_0}.
\end{eqnarray}
The Jacobian matrix $J_t$ associated with this transformation is given from Eqs.~(\ref{eq.z_XY}) and (\ref{eq.dotz_XY}):
\begin{eqnarray}
\begin{pmatrix}
z_t  \\
p_t  \\
\end{pmatrix}
=
\begin{pmatrix}
Y_t & X_t \\
\dot Y_t & \dot X_t \\
\end{pmatrix}
\begin{pmatrix}
z_{t_0}  \\
p_{t_0}  \\
\end{pmatrix}
=J_t
\begin{pmatrix}
z_{t_0}  \\
p_{t_0}  \\
\end{pmatrix},
\end{eqnarray}
which yields $|J_t|=\dot X_tY_t-X_t\dot Y_t=W_t$.

We can obtain $W_t$ in this model as follows.
By differentiating Eq.~(\ref{eq.def_W}) with respect to time and using Eqs.~(\ref{eq.X}) and (\ref{eq.Y}), we have
\begin{eqnarray}
\dot W_t=-\Gamma_t W_t.\label{eq.W_dyn}
\end{eqnarray}
By solving this equation, we obtain
\begin{eqnarray}
W_t&&=W_{t_0}\exp \left(-\int_{t_0}^t \Gamma_{t'}dt' \right)\nonumber\\
&&=\exp \left(-\int_{t_0}^t \Gamma_{t'}dt' \right)> 0, \label{eq.W_inq}
\end{eqnarray}
where we used Eq.~(\ref{eq.W_ini}) in the second equality. 

For the case $\Gamma_t=0$, we recover $W_t=1$ because the phase-plane area is conserved as seen from Eq.~(\ref{eq.W_inq}) (the Liouville's theorem).
In this case, Eq.~(\ref{eq.Q_inq}) reads~\cite{H1953}
\begin{eqnarray}
Q_t^*\ge 1,\label{eq.Q_1}
\end{eqnarray}
where the equality is achievable when the angular frequency $\omega_t$ is adiabatically changed so that $(\dot X_t^2+\omega_t^2 X_t^2)/2\omega_t$ and $(\dot Y_t^2+\omega_t^2 Y_t^2)/2\omega_t$ in Eq.~(\ref{eq.Q_def}) are conserved as the adiabatic invariants for the dynamics Eqs.~(\ref{eq.X}) and (\ref{eq.Y}) with $\Gamma_t=0$, respectively.
Thus, $Q_t^*$ measures the adiabaticity of the process.

For the case $\Gamma_t\ne 0$, it is more non-trivial to achieve the equality of Eq.~(\ref{eq.Q_inq}), and its condition is given by (see Appendix~\ref{Husimi})
\begin{eqnarray}
&&X_t=-\frac{\omega_{t_0}^{-1}}{\omega_t}\dot Y_t,\label{eq.cond_QWX}\\
&&Y_t=\frac{\omega_{t_0}}{\omega_t}\dot X_t.\label{eq.cond_QWY}
\end{eqnarray}
In the following, we show that this condition may be approximately met when a time-scale separation is assumed based on an elementary method~\cite{FGGV2002}.

We assume that the angular frequency $\omega_t$ is adiabatically (sufficiently slowly) changed in time, 
and the duration $t_1-t_0$ taken for a finite change $\omega_{t_1}-\omega_{t_0}$ of $\omega_t$ is much longer than the oscillation period $T_t\equiv 2\pi/\omega_t$ ($T_t \ll t_1-t_0$).
Moreover, we also assume that the magnitude of $\Gamma_t$ is sufficiently small ($|\Gamma_t| \ll \omega_t$).
In other words, the time constant $1/\Gamma_t$ is much larger than the oscillation period $T_t$ ($T_t \ll 1/\Gamma_t$).

Under the above assumptions, we may assume the following form for $X_t$ and $Y_t$:
\begin{eqnarray}
&&X_t=\frac{1}{\sqrt{\omega_{t_0}}}\rho_t \sin \theta_t,\label{eq.X_slow}\\
&&Y_t=\sqrt{\omega_{t_0}}\rho_t \cos \theta_t,\label{eq.Y_slow}
\end{eqnarray}
where $\rho_t$ denotes the amplitude
\begin{eqnarray}
\rho_t\equiv \sqrt{\frac{\omega_{t_0}^2X_t^2+Y_t^2}{\omega_{t_0}}},
\end{eqnarray}
which changes so slowly that it may be regarded as a constant during the oscillation period $T_t$, and $\theta_t$ denotes the phase.
By putting Eq.~(\ref{eq.X_slow}) into Eq.~(\ref{eq.X}) or Eq.~(\ref{eq.Y_slow}) into Eq.~(\ref{eq.Y}), we obtain the following same equations for $\rho_t$ and $\theta_t$:
\begin{eqnarray}
&&\ddot \rho_t-\rho_t \dot \theta_t^2=-\Gamma_t \dot \rho_t-\omega_t^2 \rho_t,\label{eq.rho}\\
&&\frac{d}{dt}\left(\rho_t^2 \dot \theta_t\right)=-\Gamma_t \rho_t^2 \dot \theta_t.\label{eq.theta}
\end{eqnarray}
Equation~(\ref{eq.rho}) with $\Gamma_t=0$ is referred to as the Ermakov equation~\cite{E1880,P1950,LA2008}.
By comparing Eqs.~(\ref{eq.W_dyn}) and (\ref{eq.theta}), we find
\begin{eqnarray}
W_t=\rho_t^2\dot \theta_t.
\end{eqnarray}
By putting $\dot \theta_t=W_t/\rho_t^2$ into Eq.~(\ref{eq.rho}), it is rewritten in terms of $W_t$ as
\begin{eqnarray}
\ddot \rho_t+\omega_t^2 \rho_t-\Gamma_t \dot \rho_t=\frac{W_t^2}{\rho_t^3},\label{eq.rho_W}
\end{eqnarray}
where $W_t$ changes slowly obeying Eq.~(\ref{eq.W_dyn}) due to the assumption of sufficiently small $|\Gamma_t|$.
Taking the time-average of Eq.~(\ref{eq.rho_W}) over the oscillation period, the terms with time derivative vanishes while during this period $\omega_t$ and $W_t$ may be regarded as time-independent constants. Therefore, we get the approximate solution to Eq.~(\ref{eq.rho_W}) as
\begin{eqnarray}
\rho_t\simeq \sqrt{\frac{W_t}{\omega_t}}.
\end{eqnarray}
Further, by using this approximate solution, we can approximate $\theta_t$ as
\begin{eqnarray}
\theta_t=\int_{t_0}^t \frac{W_{t'}}{\rho_{t'}^2}dt'\simeq \int_{t_0}^t \omega_{t'}dt'.
\end{eqnarray}
Finally, we obtain $X_t$ and $Y_t$ in Eqs.~(\ref{eq.X_slow}) and (\ref{eq.Y_slow}) as
\begin{eqnarray}
&&X_t\simeq \sqrt{\frac{W_t}{\omega_{t_0}\omega_t}}\sin \left(\int_{t_0}^t \omega_{t'}dt'\right),\\
&&Y_t\simeq \sqrt{\frac{\omega_{t_0}W_t}{\omega_t}}\cos \left(\int_{t_0}^t \omega_{t'}dt'\right),
\end{eqnarray}
respectively.
Putting these into Eq.~(\ref{eq.Q_def}) and treating $\omega_t$ and $W_t$ as if they are time-independent constants, we obtain $Q_t^*\simeq W_t$ as the equality condition of Eq.~(\ref{eq.Q_inq}).

We demonstrate the attainability of the equality in Eq.~(\ref{eq.Q_inq}) by numerically solving Eqs.~(\ref{eq.X}) and (\ref{eq.Y}).
For simplicity, we adopt the following linear protocol for changing $\omega_t$ from $\omega_{t_0}$ at $t=t_0$ to $\omega_{t_1}$ at $t=t_1$:
\begin{eqnarray}
\omega_t=\omega_{t_0}+(\omega_{t_1}-\omega_{t_0})\frac{t-t_0}{t_1-t_0} \ \ (t_0 \le t \le t_1),
\end{eqnarray}
and a time-independent friction coefficient $\Gamma_t=\Gamma$.

\begin{widetext}

\begin{figure}[!t]
\includegraphics[scale=0.4]{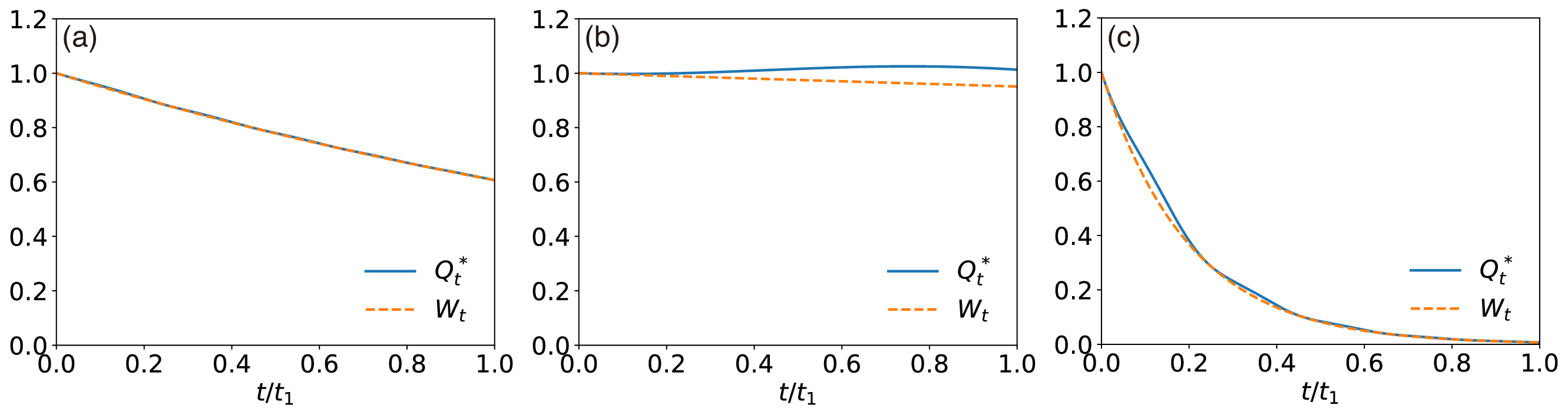}
\caption{The time evolutions of the Husimi's adiabaticity parameter $Q_t^*$ (Eq.~(\ref{eq.Q_def})) and the Wronskian $W_t$ (Eq.~(\ref{eq.def_W})) as the functions of $X_t$ and $Y_t$ obeying Eqs.~(\ref{eq.X}) and (\ref{eq.Y}), respectively, for three different $(t_1, \Gamma)$: (a) $(t_1, \Gamma)=(10, 0.05)$, (b) $(t_1, \Gamma)=(1, 0.05)$, and (c) $(t_1, \Gamma)=(10, 0.5)$. 
We used $t_0=0$, $\omega_{t_0}=1.2$, and $\omega_{t_1}=2$ as the other parameters.}\label{fig_eq_cond}
\end{figure}

\end{widetext}

In Fig.~\ref{fig_eq_cond} (a)--(c), we show the time evolutions of the Husimi's adiabaticity parameter $Q_t^*$ 
and the Wronskian $W_t$ for three different set of $(t_1, \Gamma)$. In all cases, we find that the inequality Eq.~(\ref{eq.Q_inq}) is satisfied.
In particular, in (a) where the values $(t_1, \Gamma)$ are consistent with the assumptions of the time-scale separations ($T_t \ll t_1-t_0$ and $T_t \ll 1/\Gamma_t$), we can see that $Q_t^*$ almost attains the lower bound $W_t$. Meanwhile, in (b) where only $T_t \ll 1/\Gamma_t$ is satisfied, $Q_t^*$ deviates from the bound $W_t$ because $\omega_t$ is changed fast in time, thus violating the other condition $T_t \ll t_1-t_0$. Conversely, in (c) where only $T_t \ll t_1-t_0$ is satisfied, $Q_t^*$ is relatively close to the bound $W_t$, but it is not perfect.

\subsection{The Planck principle}
Now that we have obtained the Husimi's adiabaticity parameter (Eq.~(\ref{eq.Q_def})) and the inequality for the parameter (Eq.~(\ref{eq.Q_inq})),  
we investigate how they are related to the second law of thermodynamics for the pendulum.
First, we consider the second law of thermodynamics in terms of internal energy.

For a cyclic process that lasts for $t_1-t_0$ with $\omega_{t_1}=\omega_{t_0}$ in Eq.~(\ref{eq.U}), we have
\begin{eqnarray}
E_{t_1}=E_{t_0}Q_{t_1}^*.\label{eq.Ecyc}
\end{eqnarray}
For $\Gamma_t=0$, by noting Eq.~(\ref{eq.Q_1}), we obtain
\begin{eqnarray}
E_{t_1}\ge E_{t_0},\label{eq.Planck}
\end{eqnarray}
which is Eq.~(\ref{eq.Planck_intro}) and is already known in~\cite{H1953}.
Though it is not explicitly mentioned so in~\cite{H1953}, Eq.~(\ref{eq.Planck}) may be considered as the Planck principle in thermodynamics stating that internal energy (or, equivalently, temperature) of a thermally-isolated system never decreases at the end of a cyclic process.
It should be noted that the formulation of the Planck principle in thermodynamics of small systems is not so obvious; 
see~\cite{S2002} for a counterexample to the Planck principle being true.

Meanwhile, in a general case with $\Gamma_t\ne 0$, Eq.~(\ref{eq.Ecyc}) becomes
\begin{eqnarray}
E_{t_1}\ge E_{t_0}W_{t_1},\label{eq.E_W}
\end{eqnarray}
where we used Eq.~(\ref{eq.Q_inq}).
In this case, although the internal energy may not necessarily increase depending on the dynamical factor $W_t$, 
the existence of the inequality relation is the reminiscent of the second law of thermodynamics,
which we may regard it as a generalized version of the Planck principle.

\subsection{The law of entropy non-decrease}\label{second_law}
We consider the second law of thermodynamics for the pendulum in terms of entropy, that is, the law of entropy non-decrease.
The definition of entropy which is valid for nonequilibrium processes is by no means obvious. Here, we adopt the volume entropy following~\cite{T2000,C2008_1}.
To this end, we introduce the phase-plane area enclosed by an elliptical trajectory of the pendulum with angular frequency $\omega$ and energy $E$:
\begin{eqnarray}
\Omega&&\equiv \int_{-\infty}^{\infty} \int_{-\infty}^{\infty} dz dp \Theta \left(E-H(z, p)\right)=2\pi\frac{E}{\omega},\label{eq.Omega_def}
\end{eqnarray}
where $\Theta(x)$ is a Heaviside step function satisfying $\Theta(x)=1$ for $x\ge 0$ and $\Theta(x)=0$ for $x<0$.
We also define $\Omega_t(z_t, p_t)$ as the phase-plane area enclosed by an elliptical trajectory with $\omega_t$ and $H_t$ at time $t$ as
\begin{eqnarray}
\Omega_t(z_t, p_t) \equiv 2\pi\frac{H_t(z_t, p_t)}{\omega_t},\label{eq.Omega_t_def}
\end{eqnarray}
which is conditioned on $(z_{t_0}, p_{t_0})$ as $H_t(z_t, p_t)=H_t(z_t(z_{t_0}, p_{t_0}), p_t(z_{t_0}, p_{t_0}))$ in Eq.~(\ref{eq.Omega_t_def}) is so.
Note that $\Omega_t(z_t, p_t)$ is conserved under an adiabatic change of $\omega_t$ because $H_t(z_t, p_t)/\omega_t$ serves as the adiabatic invariant (Eq.~(\ref{eq.action})).
Then, we may define the volume entropy $S_t$~\cite{T2000,C2008_1} by
\begin{eqnarray}
S_t\equiv \left<\ln \Omega_t(z_t, p_t)\right>,\label{eq.volume_S}
\end{eqnarray}
where we set the Boltzmann constant to unity and $S_{t_0}=\left<\ln \Omega_{t_0}(z_{t_0}, p_{t_0})\right>=\ln 2\pi(E_{t_0}/\omega_{t_0})$ for any initial value $(z_{t_0}, p_{t_0})$.
A related definition of entropy is a surface entropy given by the logarithm of the energy density of states. Though the volume and surface entropies for systems in equilibrium coincide in the thermodynamic limit, the volume entropy is more appropriate here for its direct connection to an adiabatic invariant.
Besides, a more appropriate choice for small systems is the volume entropy; see, e.g.,~\cite{C2008_1} and references therein including~\cite{A2004, C2005}.

With the definition Eq.~(\ref{eq.volume_S}), the entropy change is calculated as (see Appendix~\ref{DS_calc} for the derivation)
\begin{eqnarray}
\Delta S_t \equiv S_t-S_{t_0}=\left<\ln \frac{\Omega_t}{\Omega_{t_0}}\right>=\ln \frac{Q_t^*+W_t}{2}.\label{eq.DS_calc}
\end{eqnarray}
For $\Gamma_t=0$, because of $W_t=1$, Eq.~(\ref{eq.DS_calc}) recovers
\begin{eqnarray}
\Delta S_t=\ln \frac{Q_t^*+1}{2}\ge 0,\label{eq.DS_ineq}
\end{eqnarray}
where the inequality is ensured from Eq.~(\ref{eq.Q_1}) and the equality is achieved when $Q_t^*=1$.

Meanwhile, as for general cases including $\Gamma_t\ne 0$, because the area expansion rate in the phase plane $W_t$ can take values larger or smaller than unity, 
the pendulum may not be regarded as being thermally isolated.
Thus, we may not expect that we have the law of entropy non-decrease like Eq.~(\ref{eq.DS_ineq}). 
Nevertheless, by considering Eq.~(\ref{eq.Q_inq}), we notice that we can bound the volume-entropy change from the lower side as
\begin{eqnarray}
 \Delta S_t=\ln \frac{Q_t^*+W_t}{2}\ge \ln W_t.\label{eq.DS_W}
\end{eqnarray}
This inequality imposes a thermodynamic limitation on the allowed operation of the pendulum through the angular frequency even in the case of $\Gamma_t\ne 0$,
which is reminiscent of the second law of thermodynamics.
When $\Gamma_t \ne 0$, the system is no longer thermally isolated and 
we may regard that there is an implicit coupling to a surrounding such as a heat reservoir through the friction term.
However, the degrees of freedom of such a heat reservoir have not been taken into account explicitly here.
Therefore, we are using the ``second law" statement in a meaning that there is an operational restriction even on a non-Liouvillian dynamics closed only by the system's degree of freedom.
This result also highlights a role of the volume entropy in a non-thermally-isolated system beyond the role in a thermally-isolated system.

Two more remarks are in order with respect to the second laws Eqs.~(\ref{eq.DS_ineq}) and (\ref{eq.DS_W}). 
First, can we obtain the similar inequalities if we use an alternative entropy such as Gibbs entropy instead of the volume-entropy?
The Gibbs entropy is defined by
\begin{eqnarray}
S_t^{\rm G} &&\equiv \left<-\ln f_t(z_t, p_t)\right>\nonumber\\
&&=-\int_{-\infty}^\infty\int_{-\infty}^\infty dz_t dp_t \ f_t(z_t, p_t)\ln f_t(z_t, p_t).\label{eq.SH}
\end{eqnarray}
As it can be easily shown (see Appendix~\ref{Shannon} for the derivation), the Gibbs entropy change is given by
\begin{eqnarray}
\Delta S_t^{\rm G} \equiv S_t^{\rm G}-S_{t_0}^{\rm G}=\ln W_t\label{eq.DSS_W}
\end{eqnarray}
in terms only of the Wronskian $W_t$.
This does not include the Husimi's adiabaticity parameter $Q_t^*$, and the second-law-like inequalities are not obtainable.
In particular, when $\Gamma_t=0$, the Gibbs entropy is conserved for an arbitrary dynamics due to the Liouville's theorem.
This indicates that the volume entropy may be more relevant entropy to be adopted for the present model.
  
 It is noteworthy that we may express Eq.~(\ref{eq.DS_W}) in a different way in terms of the Gibbs entropy Eq.~(\ref{eq.DSS_W}):
\begin{eqnarray}
\Delta S_t \ge \Delta S_t^{\rm G} .\label{eq.DS_DSH}
\end{eqnarray}
Thus, the Gibbs entropy change may serve as a lower bound on the volume-entropy change.
Again, when $\Gamma_t=0$, the Gibbs entropy is conserved along a dynamics, and 
Eq.~(\ref{eq.DS_DSH}) consistently recovers Eq.~(\ref{eq.DS_ineq}).

Second, it is interesting to consider a relationship between the present result and the characterization of thermodynamic entropy as a Noether invariant, which is
discussed in~\cite{SY2016,BM2025}. In fact, when $\Gamma_t=0$, the following Ermakov-Lewis invariant is known as a Noether invariant of the pendulum obeying Eq.~(\ref{eq.motion})~\cite{L1967,EG1976,L1978}:
\begin{eqnarray}
I_t=\frac{1}{2}\Biggl\{\frac{z_t^2}{\rho_t^2}+(p_t \rho_t-z_t \dot \rho_t)^2\Biggr\},
\end{eqnarray}
where $\rho_t$ satisfies the Ermakov equation Eq.~(\ref{eq.rho}) with $\Gamma_t=0$.
Because the Wronskian $W_t$ is related with the Ermakov-Lewis invariant $I_t$ by $I_t=W_t^2/2$~\cite{FG2009}, we may express the entropy change in Eq.~(\ref{eq.DS_calc}) in the quasistatic limit ($Q_t^* \approx W_t=\sqrt{2I_t}$) as
\begin{eqnarray}
\Delta S_t\approx \frac{1}{2}\ln (2I_t)
\end{eqnarray}
in terms of the Noether invariant.

\section{Concluding remarks}\label{Concluding remarks}
In this paper, stimulated by the classical paper by Husimi~\cite{H1953}, we investigated the irreversibility of the damped pendulum that is initially distributed obeying the microcanonical distribution and then evolves deterministically under a time-dependent operation of the angular frequency.
We investigated the generalized version of the second law of thermodynamics for the pendulum, in terms of internal energy in Eq.~(\ref{eq.E_W}) and entropy in Eq.~(\ref{eq.DS_W}), respectively, generalizing the argument by Husimi~\cite{H1953}.
In both cases, the Husimi's adiabaticity parameter bounded by the Wronskian played a key role in quantifying the irreversibility of the thermodynamic process.

Though the present results are limited to the one-dimensional damped pendulum, 
it would be straightforward to extend the present results to multidimensional pendulums and linearly-coupled pendulums,
which are analytically tractable owing to their linearity. 
We expect that essentially the same Husimi-Wronskian framework as the present results (Eq.~(\ref{eq.Q_inq})) could be applied for such dissipative systems.
The generalization of the inequalities Eqs.~(\ref{eq.E_W}) and (\ref{eq.DS_W}) to broader classes of dissipative systems such as nonlinear pendulums beyond the linear pendulum deserves further research.
In addition, as we stated in Sec.~\ref{second_law}, our model does not include the degrees of freedom of a heat bath.
It is of importance to incorporate the degrees of freedom of a heat bath explicitly and perform the reduction of the model into the present model 
by focusing only on the system's degrees of freedom or tracing out those of the heat bath.
Deterministic thermostat models such as the Nos\'{e}-Hoover chain thermostat~\cite{MKT1992} may be appropriate models for such an investigation. In this context, the statistical-mechanics formulation of non-Hamiltonian systems~\cite{TMM1999} may be mathematically relevant and useful.
Finally, an interesting task for open systems is to reveal a role of the volume entropy in the formulation of the second law of thermodynamics for a total system consisting of a system and a heat bath.

We expect that the present results will contribute to elucidate the relationship between thermodynamics and mechanics.

\begin{acknowledgements}
The author is grateful to anonymous referees for their valuable comments on the manuscript and to Kyosuke Watanabe for careful reading of the manuscript and useful comments on it.
This work was supported by JSPS KAKENHI Grant Number 25K07163.
\end{acknowledgements}

\appendix
\section{Derivation of Eq.~(\ref{eq.U})}\label{U_deriv}
The internal energy as the average energy $E_t=\left<H_t\right>$ in Eq.~(\ref{eq.U}) is calculated as
\begin{widetext}
\begin{eqnarray}
E_t&&=\int_{-\infty}^{\infty}\int_{-\infty}^{\infty}dz_{t_0}d p_{t_0}\ \frac{\delta \left(E_{t_0}-H_{t_0}(z_{t_0}, p_{t_0})\right)}{A_{t_0}}  H_t(z_t(z_{t_0}, p_{t_0}), p_t(z_{t_0}, p_{t_0}))\nonumber\\
&&=\frac{1}{A_{t_0}}\int_{-\infty}^{\infty}\int_{-\infty}^{\infty}dz_{t_0}d p_{t_0}\  \delta \left(E_{t_0}-\frac{1}{2}(p_{t_0}^2+\omega_{t_0}^2z_{t_0}^2)\right)
\frac{1}{2}\Bigl\{p_{t_0}^2\left(\dot X_t^2+\omega_t^2X_t^2\right)+z_{t_0}^2\left(\dot Y_t^2+\omega_t^2 Y_t^2\right)+2z_{t_0}p_{t_0}\left(\dot X_t \dot Y_t+\omega_t^2X_tY_t\right)\Bigr\},\nonumber\\
\label{eq.appnd_E}
\end{eqnarray}
\end{widetext}
where we used Eqs.~(\ref{eq.E_def}), (\ref{eq.z_XY}), and (\ref{eq.dotz_XY}) in the second equality.
By performing the variable transformation ($0 \le r<\infty, 0\le \vartheta < 2\pi$):
\begin{eqnarray}
&&z_{t_0}=(r/\omega_{t_0})\cos \vartheta,\label{eq.r_theta1}\\
&&p_{t_0}=r\sin \vartheta,\label{eq.r_theta2}
\end{eqnarray}
and using the following relation 
\begin{eqnarray}
\delta \left(E_{t_0}-\frac{1}{2}(p_{t_0}^2+\omega_{t_0}^2z_{t_0}^2)\right)&&=\delta(E_{t_0}-r^2/2)\nonumber\\
&&=\frac{\delta(r-\sqrt{2E_{t_0}})}{\sqrt{2E_{t_0}}},\label{eq.delta}
\end{eqnarray}
we derive
\begin{widetext}
\begin{eqnarray}
{\rm Eq.~(\ref{eq.appnd_E})}=&&\frac{1}{A_{t_0}\omega_{t_0}\sqrt{2E_{t_0}}}\int_0^{\infty} \int_0^{2\pi}drd\vartheta\  \delta \left(r-\sqrt{2E_{t_0}}\right)r\nonumber\\
&&\times \frac{1}{2}\Bigl\{r^2 \sin^2 \vartheta \left(\dot X_t^2+\omega_t^2X_t^2\right)+\frac{r^2}{\omega_{t_0}^2} \cos^2 \vartheta \left(\dot Y_t^2+\omega_t^2 Y_t^2\right)+\frac{2r^2}{\omega_{t_0}}\cos \vartheta \sin \vartheta \left(\dot X_t \dot Y_t+\omega_t^2 X_tY_t\right)\Bigr\}\nonumber\\
&&=E_{t_0}\frac{\omega_t}{\omega_{t_0}}Q_t^*,
\end{eqnarray}
\end{widetext}
by performing the integrations, where we used Eqs.~(\ref{eq.A0}) and (\ref{eq.Q_def}) in the second equality.

\section{Derivation of Eq.~(\ref{eq.Q_inq})}\label{Husimi}
It is straightforward to derive Eq.~(\ref{eq.Q_inq}) by rewriting the following non-negative quantity~\cite{H1953}:
\begin{eqnarray}
0&& \le \omega_{t_0}\frac{\left(\omega_t X_t+\omega_{t_0}^{-1}\dot Y_t\right)^2}{2\omega_t}+\omega_{t_0}^{-1}\frac{\left(\omega_{t_0}\dot X_t-\omega_t Y_t\right)^2}{2\omega_t}\label{eq.cond_QW}\\
&&=\omega_{t_0}\frac{\left(\omega_t^2 X_t^2+2\omega_t\omega_{t_0}^{-1}X_t\dot Y_t+\omega_{t_0}^{-2}\dot Y_t^2\right)}{2\omega_t}\nonumber\\
&&+\omega_{t_0}^{-1}\frac{\left(\omega_{t_0}^2\dot X_t^2-2\omega_t\omega_{t_0}X_t\dot Y_t+\omega_t^2Y_t^2\right)}{2\omega_t}\nonumber\\
&&=\omega_{t_0}\frac{\left(\dot X_t^2+\omega_t^2X_t^2\right)}{2\omega_t}+\omega_{t_0}^{-1}\frac{\left(\dot Y_t^2+\omega_t^2Y_t^2\right)}{2\omega_t}\nonumber\\
&&+\left(X_t\dot Y_t-\dot X_tY_t\right)\nonumber\\
&&=Q_t^*-W_t.
\end{eqnarray}
Therefore, we derived $Q^*_t\ge W_t$. The equality condition is satisfied when each term of the two terms in Eq.~(\ref{eq.cond_QW}) vanishes, which yields Eqs.~(\ref{eq.cond_QWX}) and (\ref{eq.cond_QWY}).

\section{Derivation of Eq.~(\ref{eq.DS_calc})}\label{DS_calc}
The entropy change $\Delta S_t=\left<\ln \Omega_t/\Omega_{t_0}\right>$ is given as
\begin{widetext}
\begin{eqnarray}
\Delta S_t&&=\int_{-\infty}^{\infty}\int_{-\infty}^{\infty}dz_{t_0}d p_{t_0}\ \frac{\delta \left(E_{t_0}-H_{t_0}(z_{t_0}, p_{t_0})\right)}{A_{t_0}}  \ln \Biggl\{\frac{H_t(z_t(z_{t_0}, p_{t_0}), p_t(z_{t_0}, p_{t_0}))}{\omega_t} \cdot \frac{\omega_{t_0}}{E_{t_0}}\Biggr\}\nonumber\\
&&=\frac{1}{A_{t_0}}\int_{-\infty}^{\infty}\int_{-\infty}^{\infty}dz_{t_0}d p_{t_0}\  \delta \left(E_{t_0}-\frac{1}{2}(p_{t_0}^2+\omega_{t_0}^2z_{t_0}^2)\right)\nonumber\\
&&\times \ln \Biggl\{\frac{\Bigl\{p_{t_0}^2\left(\dot X_t^2+\omega_t^2X_t^2\right)+z_{t_0}^2\left(\dot Y_t^2+\omega_t^2 Y_t^2\right)+2z_{t_0}p_{t_0}\left(\dot X_t \dot Y_t+\omega_t^2 X_tY_t\right)\Bigr\}}{2\omega_t}\cdot \frac{\omega_{t_0}}{E_{t_0}}\Biggr\}. \label{eq.appnd_DS}
\end{eqnarray}
\end{widetext}
By performing the variable transformation Eqs.~(\ref{eq.r_theta1}) and (\ref{eq.r_theta2}) and using Eq.~(\ref{eq.delta}), we derive
\begin{widetext}
\begin{eqnarray}
{\rm Eq.~(\ref{eq.appnd_DS})}&&=\frac{1}{2\pi}\int_0^{2\pi} d\vartheta \ln \Biggl\{ \frac{\omega_{t_0} \sin^2 \vartheta (\dot X_t^2+\omega_t^2 X_t^2)+\omega_{t_0}^{-1} \cos^2 \vartheta (\dot Y_t^2+\omega_t^2 Y_t^2)+2\cos \vartheta \sin \vartheta (\dot X_t\dot Y_t+\omega_t^2 X_tY_t)}{\omega_t} \Biggr\}\nonumber\\
&&=\ln \frac{Q_t^*+W_t}{2}.\label{eq.deri_DS}
\end{eqnarray}
\end{widetext}
\clearpage
Here, we have calculated the integral by using the following ($a=\omega_{t_0}(\dot X_t^2+\omega_t^2 X_t^2)/\omega_t$, $b=\omega_{t_0}^{-1}(\dot Y_t^2+\omega_t^2 Y_t^2)/\omega_t$, and $c=2(\dot X_t \dot Y_t+\omega_t^2 X_tY_t)/\omega_t$):
\begin{eqnarray}
\int_0^{2\pi}&&d\vartheta \ln (a \sin^2 \vartheta+b\cos^2 \vartheta+c\cos \vartheta \sin \vartheta)\nonumber\\
&&=\int_0^{2\pi}d\vartheta \ln \left(\frac{a+b}{2}+\frac{1}{2}\sqrt{(a-b)^2+c^2}\cos (2\vartheta-\phi) \right)\nonumber\\
&&=\int_0^{2\pi}d\vartheta' \ln \left(\frac{a+b}{2}+\frac{1}{2}\sqrt{(a-b)^2+c^2}\cos \vartheta' \right)\nonumber\\
&&=2\pi \ln \left(\frac{a+b+\sqrt{4ab-c^2}}{4}\right)
\end{eqnarray}
for $4ab>c^2$, where $\phi \equiv \tan^{-1}(c/(b-a))$.
Also, we used Eq.~(\ref{eq.def_W}) in the second equality in Eq.~(\ref{eq.deri_DS}).

\section{Derivation of Eq.~(\ref{eq.DSS_W})}\label{Shannon}
From the definition of the Gibbs entropy Eq.~(\ref{eq.SH}), we can show Eq.~(\ref{eq.DSS_W}) as follows:
\begin{eqnarray}
S_t^{\rm G} &&=-\int_{-\infty}^{\infty}\int_{-\infty}^{\infty}  dz_t dp_t \ f_t(z_t, p_t)\ln f_t(z_t, p_t)\nonumber\\
&&=-\int_{-\infty}^{\infty} \int_{-\infty}^{\infty} dz_{t_0} dp_{t_0} f_{t_0}(z_{t_0}, p_{t_0})\ln \Bigl\{W_t^{-1}f_{t_0}(z_{t_0}, p_{t_0})\Bigr\}\nonumber\\
&&=-\ln W_t^{-1}+S_{t_0}^{\rm G}\nonumber\\
&&=\ln W_t+S_{t_0}^{\rm G}.
\end{eqnarray}
Here, we used $f_t(z_t,p_t)=f_{t_0}(z_{t_0},p_{t_0})|J_t|^{-1}=f_{t_0}(z_{t_0},p_{t_0})W_t^{-1}$ and $f_t(z_t, p_t)dz_tdp_{t}=f_{t_0}(z_{t_0}, p_{t_0})dz_{t_0}dp_{t_0}$ in the second equality.
It should be noted that this result is valid not only for a microcanonical distribution but for any initial distribution. 
Moreover, because the Wronskian $W_t$ may be replaced with the Jacobian $|J_t|$ of the transformation in general cases, the result is not restricted to the pendulum.


\begin{references}
\bibitem{footnote1} Hereafter, the adiabatic change of the parameter is used as the same meaning as the quasistatic change.
\end{references}
\end{document}